\documentclass[conference,twocolumn]{IEEEtran}
\usepackage{setspace}
\singlespace

\usepackage[english]{babel}
\usepackage{color}
\usepackage[final]{graphicx}
\usepackage[T1]{fontenc}
\usepackage{amsmath}
\usepackage{mathtools, cuted}
\usepackage{amsthm}
\usepackage{amstext}
\usepackage{amssymb}
\usepackage{mathrsfs}

\newcounter{example}
\newenvironment{example}[1][]{\refstepcounter{example}\par\smallskip
	\noindent \textit{Example~\theexample. #1} \rmfamily}{\smallskip}
\newtheorem{definition}{Definition}
\newtheorem{theorem}{Theorem}

\newtheorem{proposition}{Proposition}
\newtheorem{lemma}{Lemma}
\theoremstyle{remark}

\newtheorem{conjecture}{Conjecture}

\newcommand{\calP}{\mathcal{P}}

\newcommand{\calU}{\mathcal{U}}

\newcommand{\R}{{\mathbb R}}
\newcommand{\C}{{\mathbb C}}

\newcommand{\eps}{\epsilon}

\newcommand{\I}[1]{\mathrm{1}_{#1}}

\newcommand{\dif}{\textrm{d}}
\newcommand{\Supp}[1]{\mathsf{Supp}\left(#1\right)}

\newcommand{\E}[1]{{\mathbb E}\left(#1\right)}

\newcommand{\eq}[1]{\begin{equation*}
	#1
	\end{equation*}}
\newcommand{\eqn}[2]{\begin{equation}
	\label{#1}
	#2
	\end{equation}}
\newcommand{\al}[1]{\begin{align*}
	#1
	\end{align*}}

\newcommand{\by}{\times}
\newcommand{\Mat}[2]{\textnormal{M}_{#1}\left(#2\right)}

\newcommand{\bsmatrix}{\left(\begin{smallmatrix}}
	\newcommand{\esmatrix}{\end{smallmatrix}\right)}

\newcommand{\dsty}[1]{$\displaystyle #1$}
\newcommand{\ndsty}[1]{$#1$}

\IEEEoverridecommandlockouts

\allowdisplaybreaks

\begin{document}

	\title{On the Noise-Information Separation of a\\Private Principal Component Analysis Scheme}
	\author{\authorblockN{Mario Diaz\authorrefmark{1}, Shahab Asoodeh\authorrefmark{2}, Fady Alajaji\authorrefmark{3}, Tam\'{a}s Linder\authorrefmark{3}, Serban Belinschi\authorrefmark{4} and James Mingo\authorrefmark{3}}\authorblockA{\authorrefmark{1}Arizona State University and Harvard University, mdiaztor@\{asu,g.harvard\}.edu}\authorblockA{\authorrefmark{2}The University of Chicago, shahab@uchicago.edu}\authorblockA{\authorrefmark{3}Queen's University, \{fady,linder,mingo\}@queensu.ca}\authorblockA{\authorrefmark{4}Institut de Math\'{e}matiques de Toulouse, Serban.Belinschi@math.univ-toulouse.fr}}
	\maketitle

\begin{abstract}
In a survey disclosure model, we consider an additive noise privacy mechanism and study the trade-off between privacy guarantees and statistical utility. Privacy is approached from two different but complementary viewpoints: information and estimation theoretic. Motivated by the performance of principal component analysis, statistical utility is measured via the spectral gap of a certain covariance matrix. This formulation and its motivation rely on classical results from random matrix theory. We prove some properties of this statistical utility function and discuss a simple numerical method to evaluate it.
\end{abstract}

\section{Introduction}

In the last decades, privacy breaches made clear the necessity of privacy mechanisms with provable guarantees. In this context, additive noise mechanisms are a popular choice among practitioners given their ease of implementation and mathematical tractability \cite{Dwork2006}. In order to understand the trade-off between the privacy guarantees provided by and the statistical cost of this type of mechanism, it is necessary to precisely quantify privacy and statistical utility. In this paper we consider two common measures of privacy, one based on mutual information and the other one on the minimum mean-squared error (MMSE). In the context of a survey with $p$ queries and $n$ respondents, we introduce a measure of statistical utility motivated by the performance of principal component analysis (PCA), a statistical method aimed at finding the least number of variables that explain a given data set \cite[Ch.~9]{muirhead2009aspects}. More specifically, statistical utility is measured by the gap between the eigenvalues of a certain covariance matrix associated with the responses. This formulation and its motivation rely on classical results from random matrix theory. To facilitate mathematical tractability, we focus on a toy model where the eigenvalues of the data covariance matrix are either large or negligible. For this model, we derive a simple numerical method to compute the utility function. A general treatment of {\it spectrum separation} can be found in \cite[Ch.~6]{bai2010spectral}.

Private versions of PCA have been analyzed in the past, specially under the framework of differential privacy, see \cite{chaudhuri2013near} and references therein. Many of these analyses rely on results stemming from finite dimensional (random) matrix theory, see, e.g., \cite{wei2016analysis}. The approach in the present paper follows a different path, relying on asymptotic random matrix theory considerations. Our main motivation is two-fold: the behavior of the eigenvalues of certain random matrices becomes simpler when the dimensions go to infinity (Thm.~\ref{Theorem:MarchenkoPastur}) and this asymptotic behavior {\it essentially} appears in finite dimension (Thm.~\ref{Theorem:NoEigenvaluesOutsideSupport}).

In Sec.~\ref{Section:Setting} we present the setting of our problem. The statistical utility function and some of its properties are then introduced in Sec.~\ref{Section:UtilityFunction}, followed by a privacy analysis in Sec.~\ref{Section:Privacy}. In particular, we study the privacy-utility trade-off in the spirit of \cite{asoodeh2016information}, which is also related with the privacy-utility trade-offs in \cite{sankar2013utility} and references therein. In Sec.~\ref{Section:NumericalComputation} a simple numerical method for the computation of the utility function is provided. Due to space limitations, all the proofs are deferred to \cite{fullversion}.

{\it Notation.} Let $\C^+ = \{z\in\C : \Im z>0\}$ and $\C^- = - \C^+$, where $\C$ is the set of complex numbers and $\Im z$ is the imaginary part of $z\in\C$. For a $p\by p$ complex matrix $A\in\Mat{p}{\C}$, we let $A_{ij}$ be its $i,j$-entry and $A^*$ be its conjugate transpose. The indicator function of a set $E$ is denoted by $\I{E}$. For a probability distribution $F$, we let $\Supp{F}$ be its support, i.e., the smallest closed set $E$ with $F(E)=1$. For $A\in\Mat{p}{\C}$ with eigenvalues $\lambda_1,\ldots,\lambda_p$, the probability distribution defined by \ndsty{F_A([a,b]) = \frac{1}{p} \sum_{k} \I{\lambda_k\in[a,b]}} is called the eigenvalue distribution of $A$.

\section{Setting}
\label{Section:Setting}

Assume that a survey with $p$ queries is handed to $n$ respondents. Let $\widehat{X}$ be the $p\by n$ matrix associated to this survey. We assume that $\widehat{X}$ is a realization of a random matrix
\eq{X = \Sigma^{1/2} W,}
where $\Sigma$ is a $p\by p$ (deterministic) covariance matrix and $W$ is a $p\by n$ random matrix whose entries are independent and identically distributed (i.i.d.) real random variables with zero mean and unit variance. Note that the columns of $\widehat{X}$ are independent realizations of a random vector with covariance $\Sigma$. A popular instance of this model corresponds to the case where the entries of $W$ are i.i.d.\ Gaussian random variables; thus the entries of $X$ are possibly correlated Gaussian random variables. The covariance matrix $\Sigma$ possesses valuable statistical information about the respondent population. Hence, in many applications the data aggregator is interested in obtaining an estimation of $\Sigma$. In this setting, the canonical estimator is the sample covariance matrix
\eq{\widehat{\Sigma} := \frac{1}{n} \widehat{X}\widehat{X}^*.}

Because of privacy concerns, the respondents might not want to disclose their answers, $\widehat{X}$, to the data aggregator. Instead, they might want to use a randomized mechanism to alter their answers, giving them the position of plausible deniability towards their responses. In this paper we focus on an additive noise model: instead of providing $\widehat{X}$ to the data aggregator, the respondents provide
\eq{\widehat{X}_t := \widehat{X} + \sqrt{t} \widehat{Z},}
where $t>0$ is a design parameter and $\widehat{Z}$ is a realization of $Z$, a $p\by n$ random matrix which is independent of $X$ and whose entries are i.i.d.\ random variables with zero mean and unit variance. In this case, the sample covariance matrix equals
\eq{\widehat{\Sigma}_t = \frac{1}{n} \widehat{X}_t \widehat{X}_t^*,}
a realization of $\Sigma_t = n^{-1} X_t X_t^*$, where $X_t := X + \sqrt{t} Z$. Note that $\widehat{\Sigma}_0 = \widehat{\Sigma}$ and that $\E{\Sigma_t} = \Sigma + t {\rm I}_p$, where ${\rm I}_p$ denotes the $p\by p$ identity matrix. The probability distribution of the additive noise may change according to the nature of the data, e.g., discrete or continuous. In particular, both $t$ and the distribution of the noise are the design parameters of the privacy mechanism. Observe that given these parameters, this additive mechanism can be implemented locally at each user, making unnecessary the presence of a trustworthy data aggregator.

If for the application at hand the noise distribution is fixed, then the trade-off between privacy and statistical utility becomes evident: when $t$ increases the respondents's privacy improves as their answers get more distorted but, at the same time, the sample covariance matrix $\widehat{\Sigma}_t$ differs more from $\Sigma$. Note that for $p$ fixed and $n$ large ($n\to\infty$), the latter is not a problem. Indeed, under some mild assumptions, the law of large numbers implies
\eq{\lim_{n\to\infty} \| \Sigma_t - (\Sigma + t {\rm I}_p) \|_2^2 \stackrel{\text{a.s.}}{=} 0,}
where a.s. stands for almost surely. Hence, the data aggregator might use $\widehat{\Sigma}_t - t {\rm I}_p$ as an estimate of $\Sigma$ without incurring a big statistical loss. However, when both $p$ and $n$ are large ($p,n\to\infty$), the estimator $\widehat{\Sigma}_t - t {\rm I}_p$ is known to be a poor estimate of $\Sigma$, e.g., the eigenvalues of $\widehat{\Sigma}_t - t {\rm I}_p$ might be very different from those of $\Sigma$, see Thm.~\ref{Theorem:MarchenkoPastur}. Since in many contemporary applications $p$ and $n$ are within the same order of magnitude, it is necessary to quantify the statistical cost incurred by the additive noise mechanism in this regime. In the next section we do so by introducing a utility function connected to the performance of PCA.

\section{Statistical Utility Function}
\label{Section:UtilityFunction}

We now introduce a statistical utility function that captures the performance of PCA applied to $\widehat{\Sigma}_t$. In order to motivate its definition, let us consider the following example.

To simplify the exposition, in this section we assume that the entries of $W$ and $Z$ are Gaussian. At the end of this section we comment on the universality of the subsequent analysis.

\begin{figure}[t]
\centering
\includegraphics[width=0.18\textwidth]{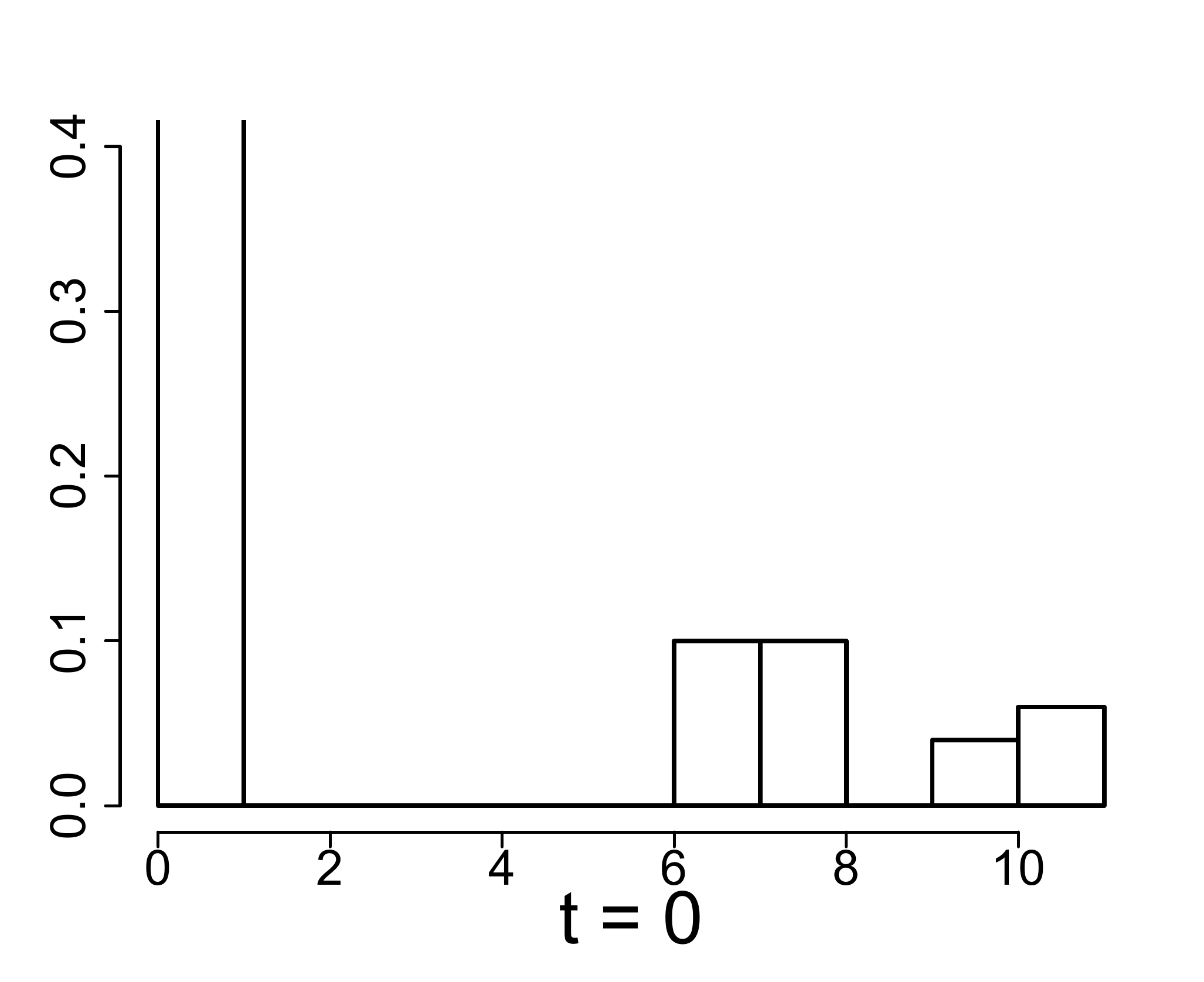}\quad\includegraphics[width=0.18\textwidth]{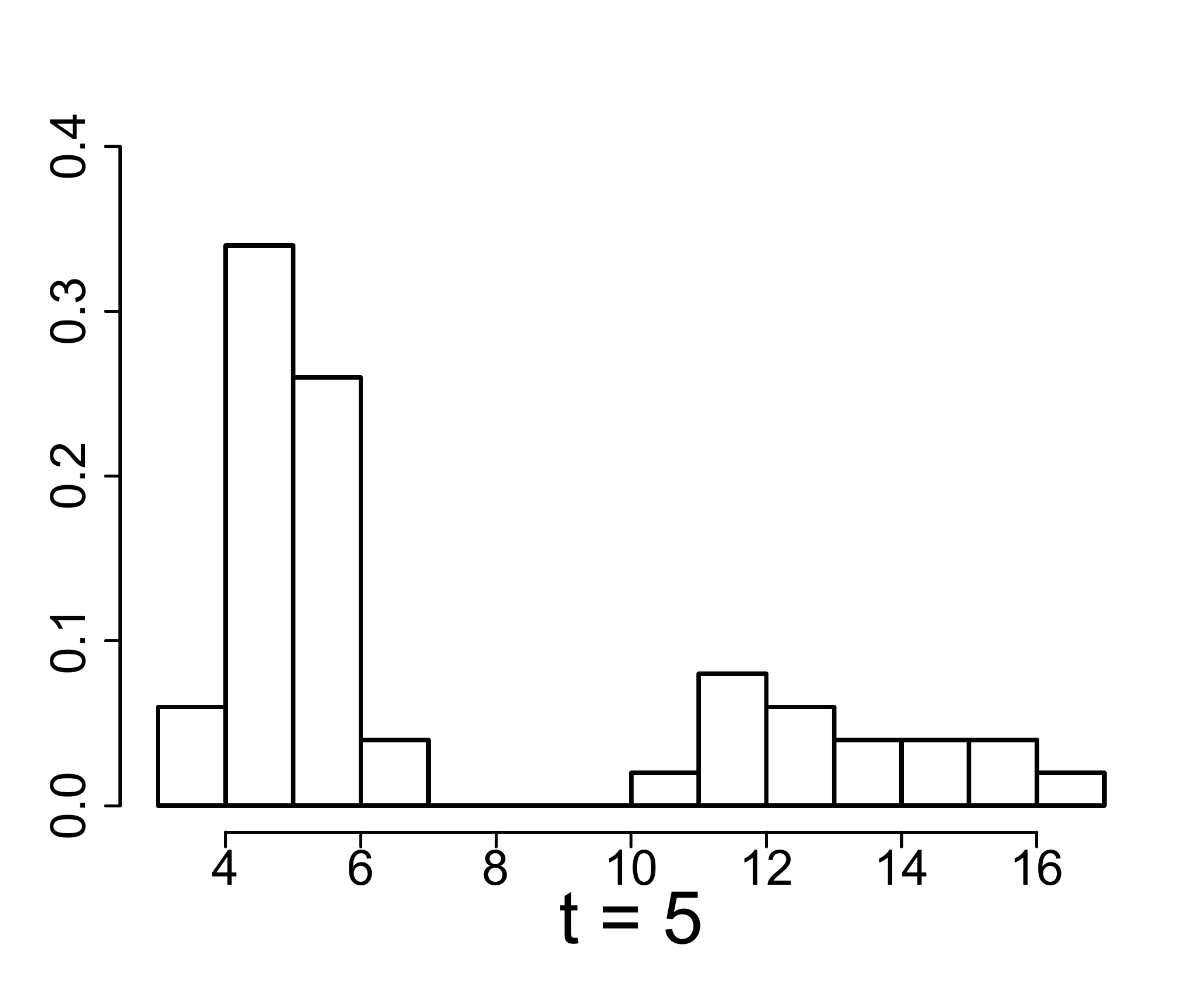}

\vspace{-10pt}

\includegraphics[width=0.18\textwidth]{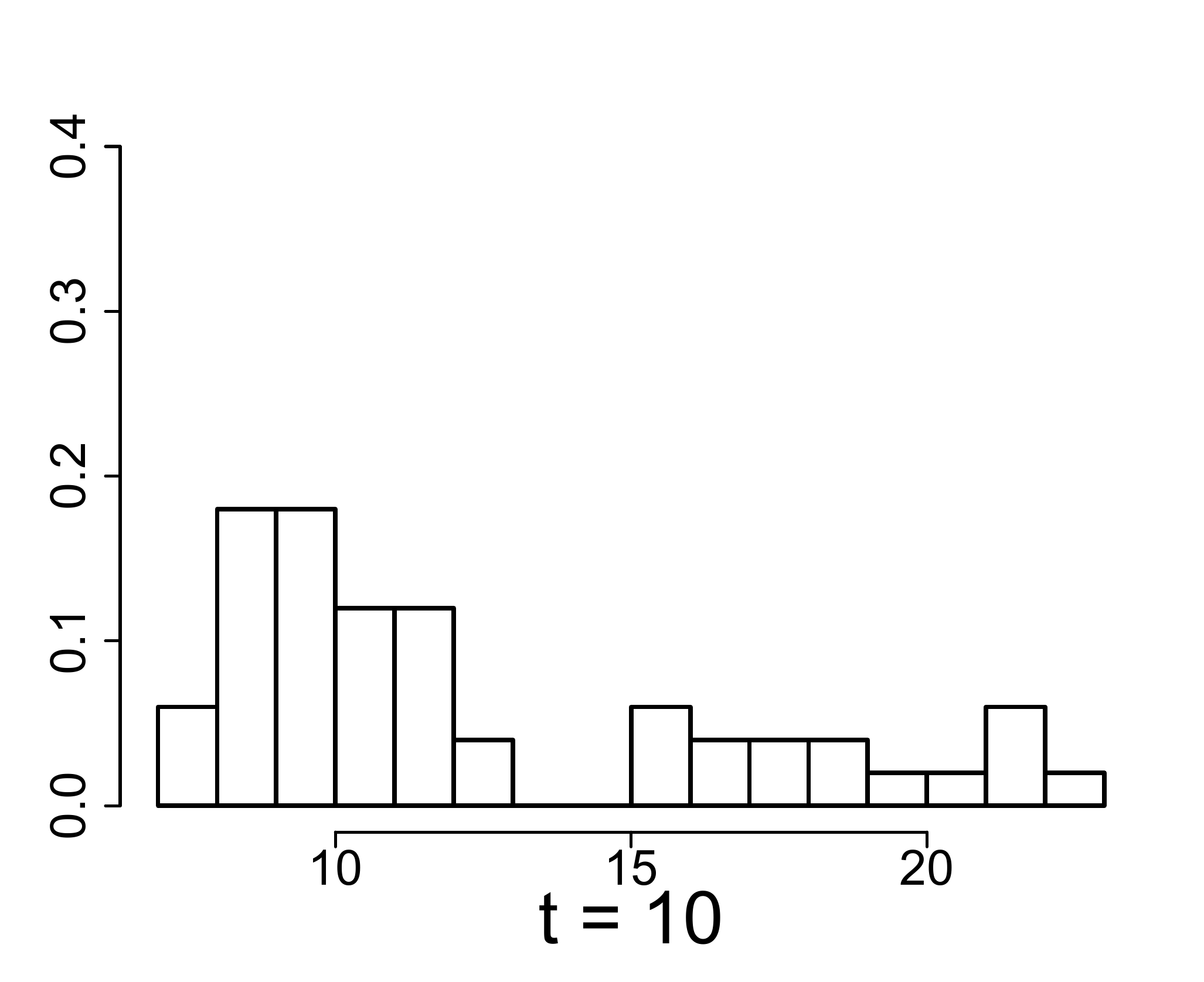}\quad\includegraphics[width=0.18\textwidth]{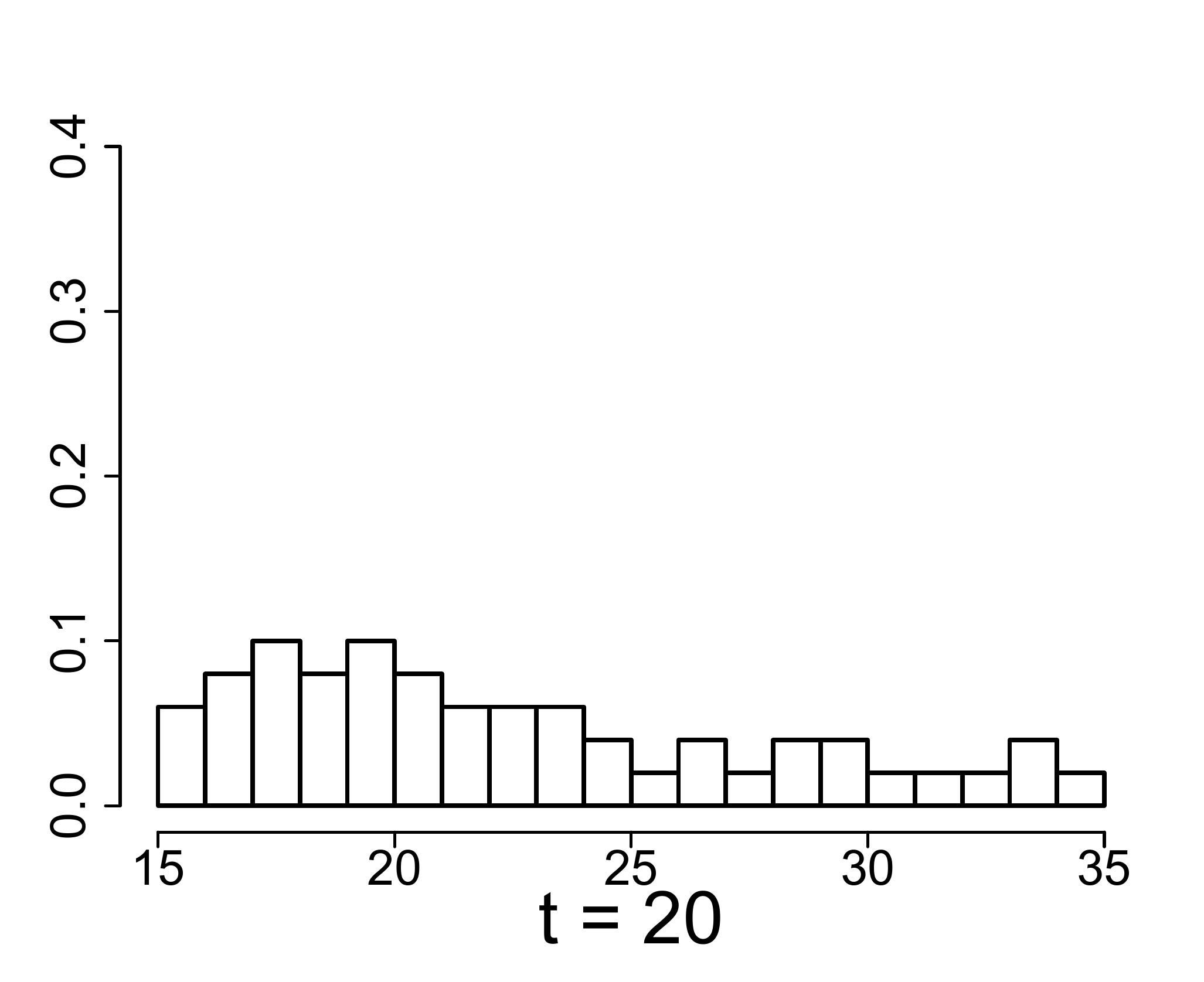}
\caption{Histogram of the eigenvalues of $\widehat{\Sigma}_t$ for different values of $t$.}
\label{Fig:ts}
\vspace{-10pt}
\end{figure}

\begin{example}
Let $p=50$ and $n=2000$. Assume that $\Sigma$ is diagonal with eigenvalues $0$, $7$, and $10$ with multiplicities $35$, $10$, and $5$, respectively. A histogram of the eigenvalues of an instance of $\widehat{\Sigma}=\widehat{\Sigma}_0$ is given in Fig.~\ref{Fig:ts}. Note that this distribution is a {\it blurred} version of the eigenvalue distribution of $\Sigma$,
\eqn{eq:FSigma}{F_{\Sigma}([a,b]) = \frac{7}{10} \I{0\in[a,b]} + \frac{2}{10} \I{7\in[a,b]} + \frac{1}{10} \I{10\in[a,b]}.}
As $t$ increases, the additive noise $\sqrt{t} \widehat{Z}$ becomes stronger, making the eigenvalue distribution of $\widehat{\Sigma}_t$ more diffuse, as shown in Fig.~\ref{Fig:ts}. This behavior has a direct impact on the benefits of PCA, which provides a dimensionality reduction inversely proportional to the number of largest eigenvalues. For example, PCA performed on $\widehat{\Sigma}=\widehat{\Sigma}_0$ would propose the five largest eigenvalues as the most informative components. Similarly, PCA performed on $\widehat{\Sigma}_5$ or $\widehat{\Sigma}_{10}$ would suggest the fifteen largest eigenvalues. Since all the eigenvalues of  $\widehat{\Sigma}_{20}$ are merged together, PCA in this latter case might be ineffective.
\end{example}

The forthcoming definition of the statistical utility function $\calU$ relies on the following asymptotic considerations. The Gaussianity assumed in this section implies
\eqn{eq:DistributionXt}{X_t = \Sigma^{1/2} W + \sqrt{t} Z \stackrel{\text{d}}{=} (\Sigma+t {\rm I}_p)^{1/2} W,}
where $\stackrel{\text{d}}{=}$ stands for equality in distribution. In particular,
\eqn{eq:DistributionSigmat}{\Sigma_t \stackrel{\text{d}}{=} \frac{1}{n} (\Sigma+t {\rm I}_p)^{1/2} WW^* (\Sigma+t {\rm I}_p)^{1/2}.}
The next theorem \cite[Thm.~1.1]{silverstein1995strong} is a generalization of the Marchenko-Pastur theorem, a cornerstone of random matrix theory. For a probability distribution function $F$, its Cauchy transform $G:\C^+\to\C^-$ is the (analytic) function defined by
\eq{G(z) = \int_\R \frac{1}{z-x} \dif F(x).}
The Cauchy transform characterizes a distribution function. Indeed, the Stieltjes inversion formula states that
\eq{F([a,b]) = -\frac{1}{\pi} \lim_{\eps\to0^+} \int_a^b \Im G(x+i\eps) \dif x,}
for all $a<b$ continuity points of $F$. When $F$ is regular enough, its density equals \dsty{f(x) = - \frac{1}{\pi} \lim_{\eps\to0^+} \Im G(x+i\eps)}.

\begin{theorem}[\!\!\cite{silverstein1995strong}]
\label{Theorem:MarchenkoPastur}
Assume on a common probability space:

\noindent(a) For $p=1,2,\ldots$, $W_p=(W_p(i,j))$ is $p\by n$, $W_p(i,j)\in\C$ are identically distributed for all $p,i,j$, independent across $i,j$ for each $p$, $\E{|W_1(1,1) - \E{W_1(1,1)}|^2}=1$;

\noindent(b) $n=n(p)$ with $p/n\to c\in(0,\infty)$ as $p\to\infty$;

\noindent(c) $T_p$ is $p\by p$ random Hermitian nonnegative definite, with eigenvalue distribution converging a.s. in distribution to a probability distribution $H$ on $[0,\infty)$ as $p\to\infty$;

\noindent(d) $W_p$ and $T_p$ are independent.

Let $T_p^{1/2}$ be the Hermitian nonnegative square root of $T_p$, and let $\Sigma_p = (1/n) T_p^{1/2} W_pW_p^* T_p^{1/2}$. Then, a.s., the eigenvalue distribution of $\Sigma_p$ converges in distribution, as $p\to\infty$, to a non-random probability distribution $F$, whose Cauchy transform $G(z)$ satisfies
\eqn{eq:MP}{G(z) = \int_\R \frac{1}{z-x(1-c+czG(z))} \dif H(x),}
in the sense that, for each $z\in\C^+$, $G(z)$ is the unique solution to \eqref{eq:MP} in $D_{c,z} = \{G\in\C : (1-c)/z+cG\in\C^-\}$.
\end{theorem}

Note that if $H$ is discrete, the integral in \eqref{eq:MP} reduces to a sum. In particular, if $|\Supp{H}\cap(0,\infty)|=n$, then $G(z)$ is the only root in $D_{c,z}$ of a polynomial of degree $n+1$. For instance, if $H(x) = \I{x\geq1}$, then $G=G(z)$ solves the equation \dsty{czG^2 + (1-c-z)G +1 = 0}, a quadratic polynomial in $G$. The next example shows the predictive power of Thm.~\ref{Theorem:MarchenkoPastur}.

\begin{figure}[t]
\centering
\includegraphics[width=0.20\textwidth]{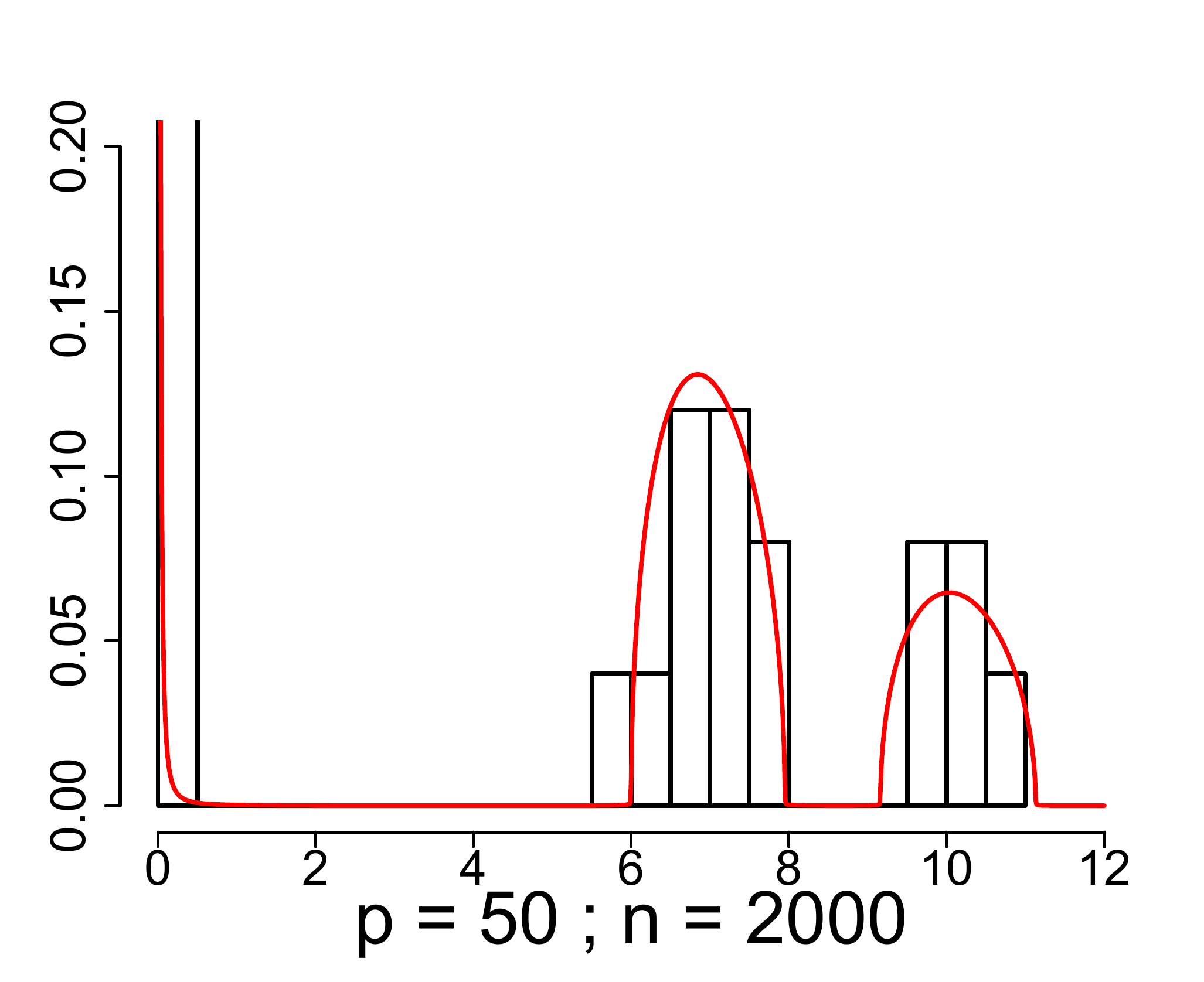}\quad\includegraphics[width=0.20\textwidth]{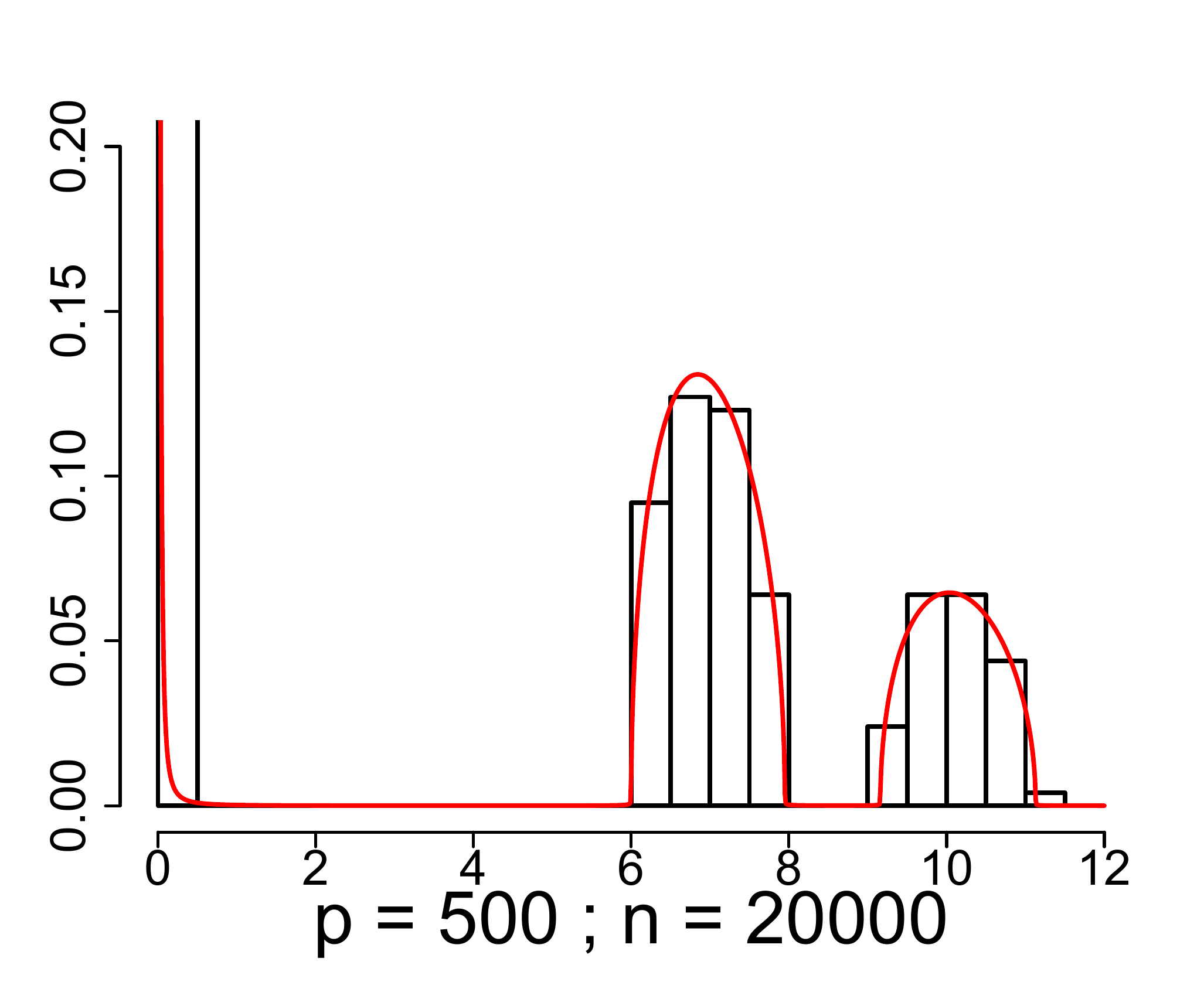}
\caption{Histogram of the eigenvalues of $\widehat{\Sigma} = \widehat{\Sigma}_0$ and the asymptotic density predicted by Thm.~\ref{Theorem:MarchenkoPastur}.}
\label{Fig:Prediction}
\vspace{-10pt}
\end{figure}

\begin{example}
In Fig.~\ref{Fig:Prediction}, the histogram of the eigenvalues of a realization of $\widehat{\Sigma}_0$ is depicted for two different values of $p$ and $n$ with $c=1/40$. In both cases, the eigenvalue distribution of $\Sigma$ is given by \eqref{eq:FSigma}. The asymptotic density of the eigenvalues provided by Thm.~\ref{Theorem:MarchenkoPastur} is also depicted. Observe the close agreement between the empirical and asymptotic eigenvalue distributions, even for $p$ as small as 50.
\end{example}

The previous example demonstrates that not only the distribution of the eigenvalues follows closely the corresponding asymptotic density, but also that there are no eigenvalues outside the support of the asymptotic prediction. This observation is formalized in the following theorem \cite{bai1998no}. Given $c\in(0,\infty)$ and $H$ a probability distribution on $[0,\infty)$, we let $F^{c,H}$ be the limiting distribution determined by \eqref{eq:MP}.

\begin{theorem}[\!\!\cite{bai1998no}]
\label{Theorem:NoEigenvaluesOutsideSupport}
Assume:

\noindent(a) $W(i,j)\in\C$, $i,j=1,2,\ldots$, are i.i.d. random variables with $\E{W(1,1)}=0$, $\E{|W(1,1)|^2}=1$, and $\E{|W(1,1)|^4}<\infty$;

\noindent(b) $n=n(p)$ with $c_p = p/n \to c\in(0,\infty)$ as $p\to\infty$;

\noindent(c) For each $p$, $T_p$ is $p\by p$ Hermitian nonnegative definite with eigenvalue distribution $H_p$ converging in distribution to a probability distribution $H$;

\noindent(d) $\Sigma_p = (1/n) T_p^{1/2} W_pW_p^* T_p^{1/2}$ where $W_p=(W_p(i,j))$ with $i=1,2,\ldots,p$, $j=1,2,\ldots,n$ and $T_p^{1/2}$ is any Hermitian square root of $T_p$;

\noindent(e) The interval $[a,b]$ with $a>0$ lies outside the support of $F^{c,H}$ and $F^{c_p,H_p}$ for all large $p$.

Then, with probability one, no eigenvalue of $\Sigma_p$ appears in $[a,b]$ for all large $p$.
\end{theorem}

The previous theorem readily implies that {\it the gaps in the support of $F^{c,H}$ appear in finite dimension}. This is of particular interest for this paper, as PCA is more useful when there are few large eigenvalues, i.e., there is a gap between large and small eigenvalues. Now we introduce the promised statistical utility function.

In order to keep the analysis tractable, we consider the following toy model for the situation in which there is a clear distinction between large and small eigenvalues: {\it the covariance matrix $\Sigma$ has only one non-zero eigenvalue, say $s$, with multiplicity $\lfloor rp \rfloor$ for some $r\in(0,1)$.} Under this assumption, the eigenvalue distribution of $\Sigma+t{\rm I}_p$ equals\footnote{More precisely, $\| H_t - F_{\Sigma+t{\rm I}_p}\|_\infty \leq 1/p$ which is negligible.}
\eqn{eq:Ht}{H_t(x) = (1-r) \I{x\geq t} + r \I{x\geq s+t}.}
By \eqref{eq:DistributionSigmat} and Thm.~\ref{Theorem:MarchenkoPastur}, a.s., the eigenvalue distribution of $\Sigma_t$ converges, as $p\to\infty$, to
\eqn{eq:Ft}{F_t := F^{c,H_t}.}
Finally, let $\Supp{F_t}$ be the support of $F_t$ and $N(t)$ be the number of its connected components. Note that, by Lemma~\ref{Lemma:Delta}, $N(t)$ is finite for every $t\in[0,\infty)$.

\begin{definition}
The utility function $\calU:[0,\infty)\to\R$ is defined as follows. If $N(t) = 1$, we let $\calU(t)=0$. If $N(t)=2$ and $A_t,B_t$ are the connected components of $\Supp{F_t}$,
\eq{\calU(t) = \min_{a\in A_t, b\in B_t} |a-b|.}
\end{definition}

In words, $\calU(t)$ approximates the separation between the large and the small eigenvalues of $\widehat{\Sigma}_t$, as long as such separation exists. As exhibited by equations \eqref{eq:DistributionXt} and \eqref{eq:DistributionSigmat}, the large eigenvalues of $\widehat{\Sigma}_t$ correspond mainly to the non-zero eigenvalues of $\Sigma$, while the small ones come from the added noise $\sqrt{t}Z$. Note that in order for $\calU$ to be well defined, it is necessary for the range of $N$ to be a subset of $\{1,2\}$, as established next.

\begin{theorem}
\label{Theorem:NRange}
With the assumptions from \eqref{eq:Ht} and \eqref{eq:Ft}, we have that $N(t)\in\{1,2\}$ for all $t\in[0,\infty)$.
\end{theorem}

\begin{figure}[t]
\centering
\includegraphics[width=0.30\textwidth]{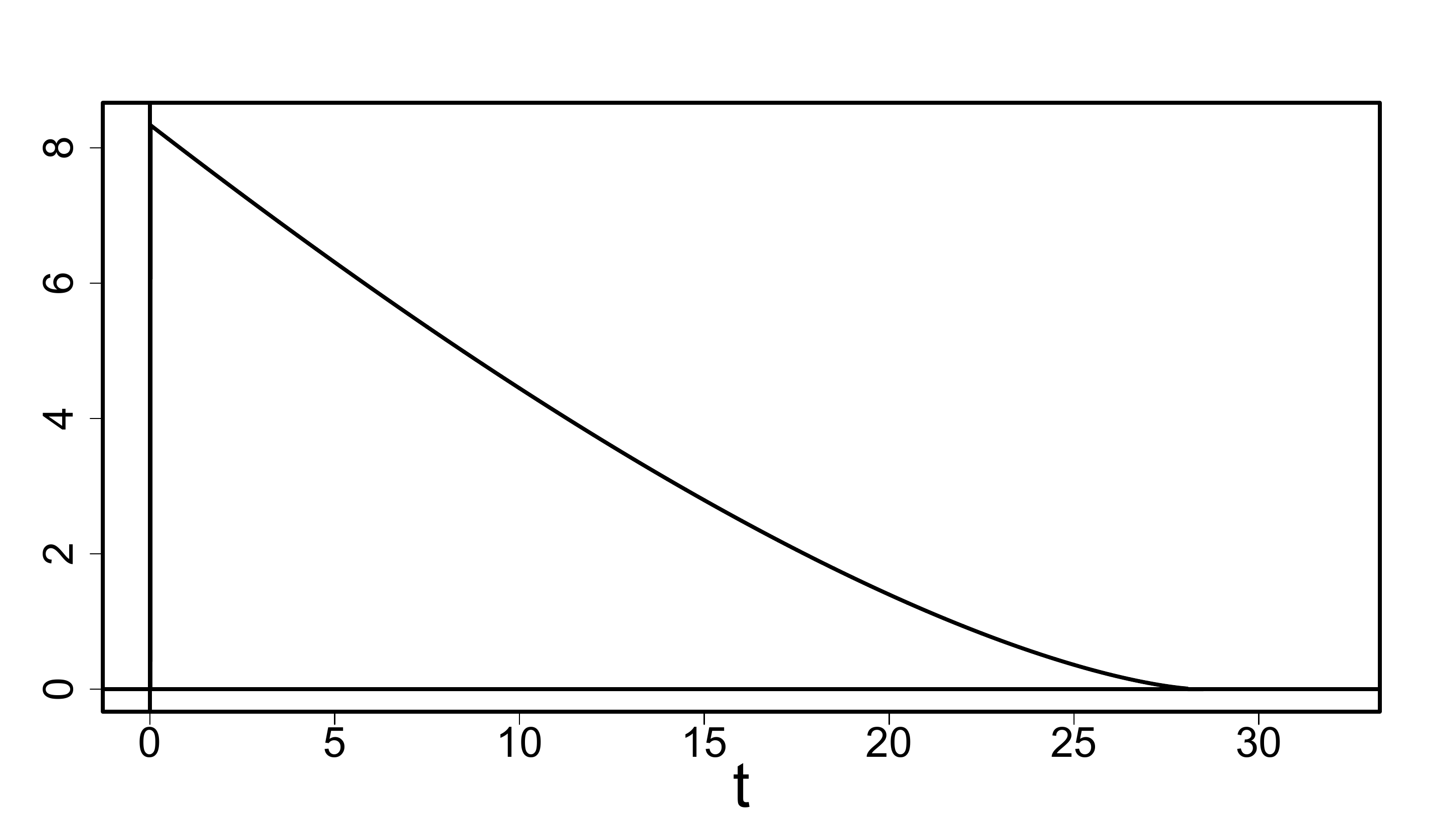}
\caption{The graph of $\calU$ for $c=1/40$, $r=3/10$, and $s=10$.}
\label{Fig:GraphU}
\vspace{-10pt}
\end{figure}

One way to compute $\calU(t)$ is finding $F_t$, determining its connected components, and measuring their distance. However, there is a more efficient method based on the discriminant of a cubic equation. To avoid an unnecessary digression, this method is discussed in Sec.~\ref{Section:NumericalComputation}. Using this method, in Fig.~\ref{Fig:GraphU} we plot the graph of $\calU$.

Under our standing assumptions, the performance of PCA is heavily compromised for a noise power $t$ such that $\calU(t)=0$, as the gap between noise and information disappears. Indeed, for $t$ large enough the gap always disappears, as established by the following proposition.

\begin{proposition}
\label{Proposition:NAsymptotic}
There exists $T=T(c,r,s)\geq0$ such that $N(t)=1$ for all $t\geq T$.
\end{proposition}

Note that in Fig.~\ref{Fig:GraphU} there exists a $t^*$ such that $\calU(t)=0$ if and only if $t\geq t^*$. Thus, in principle, any noise power $t\in[0,t^*]$ does not compromise the performance of PCA. This property makes $t^*$ useful in the design of privacy mechanisms. In view of Thm.~\ref{Theorem:NRange} and Prop.~\ref{Proposition:NAsymptotic}, the existence of such $t^*$ is equivalent to the following.

\begin{conjecture}
\label{Conjecture:NNonIncreasing}
$N$ is non-increasing in $t$.
\end{conjecture}

In addition to simulations, there are theoretical reasons to believe in the above conjecture, e.g., similar results are known to be true for other random matrix models \cite{biane1997free}. Ultimately, we are interested in the statistical utility function $\calU$ and not only in the set $\{t\geq0 : \calU(t)=0\} = \{t\geq0 : N(t)=1\}$. For this utility function, there is numerical evidence supporting the following stronger conjecture.

\begin{conjecture}
\label{Conjecture:UNonIncreasing}
$\calU$ is non-increasing and convex in $t$.
\end{conjecture}

\noindent{\bf Remark.} In this section we assumed that both data and noise are Gaussian. Nonetheless, one can appeal to {\it universality} arguments to establish that the conclusions reached in this section hold for a much wider range of random matrix models. In the square case, when $p=n$, one can appeal to the universality of the {\it circular law}, as established in \cite{tao2010random}, and the {\it asymptotic freeness} of several random matrices, see, e.g., \cite{nica2006lectures,mingo2017free}. The non-square case can be handled similarly using the ideas in \cite{benaych2009rectangular} and references therein.

\section{Privacy Measures}
\label{Section:Privacy}

Having defined $\mathcal{U}$ as the utility function, we need to specify a privacy function to quantify the trade-off between utility and privacy. A natural option is to measure the information leakage of the user's raw data in its perturbed version. In this section we discuss two specific measures of information leakage: mutual information and MMSE.

{\it Mutual Information.} Let $X^{(j)}$ and $X_t^{(j)}$, $j\leq n$, denote the $j$-th column of $X$ and $X_t$, respectively. Since the entries of $W$ and $Z$ are i.i.d., the mutual information $I(X^{(j)}; X^{(j)}_t)$ does not depend on $j$. Thus, w.l.o.g., we define
$$\mathcal{P}_{\mathsf{IT}}(t):= I(X^{(1)}; X^{(1)}_t),$$
as a privacy measure. Measuring privacy in terms of mutual information has been explored extensively in the past, see, e.g., \cite{calmon2017principal}. Assuming that both data and noise are Gaussian,
\eq{\calP_{\mathsf{IT}}^{\mathsf{G}}(t) = \frac{1}{2} \log \det \left({\rm I}_p + \frac{\Sigma}{t}\right).}
In particular, for the toy model of the previous section,
\eq{\calP_{\mathsf{IT}}^{r,s,p}(t) = \frac{\lfloor rp \rfloor}{2} \log\left(1+\frac{s}{t}\right).}
In the context of the last remark of the previous section, i.e., when data and/or noise are not necessarily Gaussian, it is relevant to consider the following.

Assume that the noise is Gaussian but the data is drawn from an arbitrary distribution having a density and finite third moment. Let $\theta=\frac{1}{\sqrt{t}}$. With this notation,
\eq{\calP_{\mathsf{IT}}(t) = h\left(\theta X^{(1)}+Z^{(1)}\right)-\frac{p}{2}\log 2\pi e,}
where $h(\cdot)$ denotes differential entropy. In particular, studying $t \mapsto \calP_{\mathsf{IT}}(t)$ amounts to studying $\theta \mapsto h(\theta X^{(1)}+Z^{(1)})$. If $p=1$, then it follows from \cite[Lemma 1]{Sensitivity} that, as $\theta\to0$, 
$$I(X^{(1)}; \theta X^{(1)}+Z^{(1)})=\frac{\theta^2}{2} + o(\theta^2),$$ 
and thus $\calP_{\mathsf{IT}}(t) \sim \frac{1}{2t}$ in the high privacy regime ($t\to \infty$). For $p\geq 1$, the chain rule implies
\eq{I(X^{(1)}; \theta X^{(1)}+Z^{(1)}) \leq  \frac{p \mathsf{Tr}(\Sigma) \theta^2}{2} + o(\theta^2),}
and hence $\calP_{\mathsf{IT}}(t) \lesssim \frac{p}{2t}\mathsf{Tr}(\Sigma)$ in the high privacy regime.

Now assume that neither data nor noise is Gaussian. Recall that the non-Gaussianity $D(V)$ of a random vector $V$ is defined as $D(V)\coloneqq D(V\|V_\mathsf{G})$, where $D(\cdot \| \cdot)$ denotes the Kullback-Leibler divergence, and $V_\mathsf{G}$ is a Gaussian random vector with the same mean and covariance matrix as $V$. It can be shown that
$$I(X^{(1)}; X_t^{(1)}) = \mathcal{P}^\mathsf{G}_{\mathsf{IT}}(t) +D(\sqrt{t}Z^{(1)})-D(X_t^{(1)}).$$
In this case, regardless of distributions of $X$ and $Z$,
$$\mathcal{P}_{\mathsf{IT}}(t)\leq \mathcal{P}^\mathsf{G}_{\mathsf{IT}}(t) + D(\sqrt{t}Z^{(1)}) = \mathcal{P}^\mathsf{G}_{\mathsf{IT}}(t) + pD(\sqrt{t}Z_{11}),$$
where the last equality holds as the entries of $Z$ are i.i.d.

{\it MMSE.} In \cite{Asoodeh_MMSE}, see also \cite{calmon2017principal}, the authors proposed an estimation-theoretic measure in terms of MMSE. Following this approach, we define
\eq{\mathcal{P}_{\mathsf{ET}}(t) \coloneqq\sum_{i=1}^p\mathsf{mmse}(X_{i1}|X_t) =\mathbb{E}\left[\big\|X^{(1)}-\mathbb{E}[X^{(1)}|X_t]\big\|^2\right],}
where $\mathsf{mmse}(U|V):= \mathbb{E}[(U-\mathbb{E}[U|V])^2]$. If both data and noise are Gaussian, then we can write
$$\mathcal{P}_{\mathsf{ET}}^{\mathsf{G}}(t)=\mathsf{Tr}\left[\left({\rm I}_p+t^{-1}\Sigma\right)^{-1}\Sigma\right].$$
In particular, for the toy model in the previous section
\eq{\mathcal{P}_{\mathsf{ET}}^{\mathsf{r,s,p}}(t) = \lfloor rp \rfloor \frac{ts}{t+s}.}
It is worth pointing out that $\mathcal{P}_{\mathsf{ET}}(t)$ and $\mathcal{P}_{\mathsf{IT}}(t)$ are connected by the so-called I-MMSE relation, see \cite{MMSE_Guo}. For example, when the noise is Gaussian, \dsty{\mathcal{P}_{\mathsf{ET}}(t) = - (2t^2) \mathcal{P}_{\mathsf{IT}}'(t)}. In this case, $\calP_\mathsf{ET}$ quantifies the rate of decrease of the information-theoretic privacy leakage $\calP_{\mathsf{IT}}$.

\begin{figure}[t]
\centering
\includegraphics[width=0.30\textwidth]{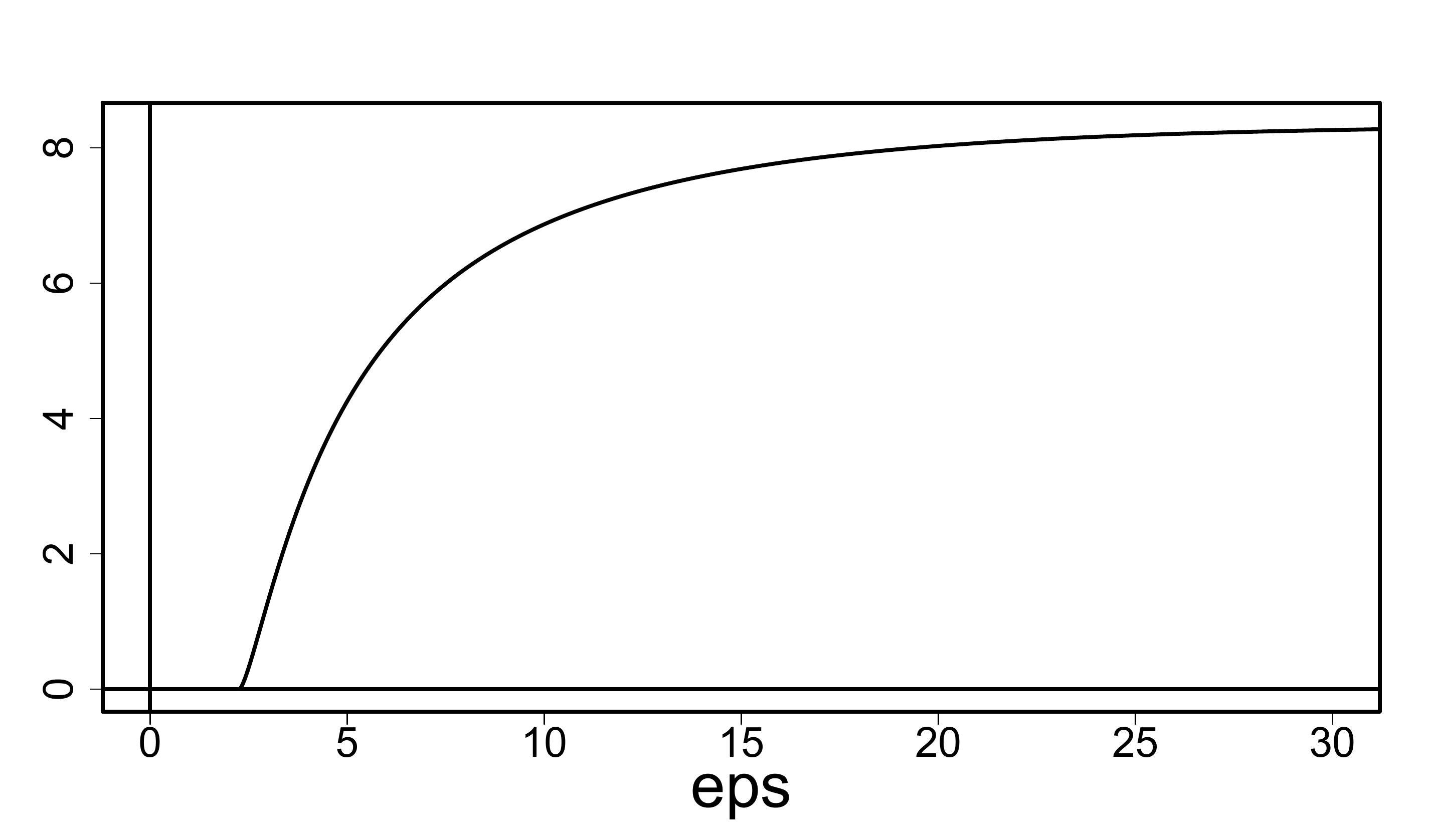}
\caption{The graph of $g_{\mathsf{IT}}^{r,s,p}$ for $r=3/10$, $s=10$ and $p=50$.}
\label{Fig:Graphg}
\vspace{-10pt}
\end{figure}

{\it Privacy-Utility Function.} In order to formally connect privacy and utility, we define the following privacy-utlity function in the spirit of \cite{asoodeh2016information}, see also \cite{sankar2013utility} and references therein. For $\epsilon>0$, we define
\eq{g_{\mathsf{IT}}(\epsilon) := \sup_{t : \calP_{\mathsf{IT}}(t) \leq \eps} \calU(t).}
In words, $g_{\mathsf{IT}}(\epsilon)$ equals the largest utility $\calU(t)$ under the privacy constraint $\calP_{\mathsf{IT}}(t) \leq \eps$. Conditional on the non-increasing behavior of $\calU(t)$ (Conj.~\ref{Conjecture:UNonIncreasing}), it is easy to verify that for the model of the previous section
\eqn{eq:PUFunctionIT}{g_{\mathsf{IT}}^{r,s,p}(\epsilon) = \calU((\calP_{\mathsf{IT}}^{r,s,p})^{-1}(\epsilon)),}
where \dsty{(\calP_{\mathsf{IT}}^{r,s,p})^{-1}(\epsilon) = s(e^{2\epsilon/\lfloor rp \rfloor}-1)^{-1}}. Since $\calU(t)$ can be computed using the tools from the following section, \eqref{eq:PUFunctionIT} provides a useful way to compute the privacy-utility function $g_{\mathsf{IT}}^{r,s,p}$. Fig.~\ref{Fig:Graphg} depicts $g_{\mathsf{IT}}^{r,s,p}$ for $r=3/10$, $s=10$ and $p=50$. Observe that, conditional on the existence of $t^*$ (Conj.~\ref{Conjecture:NNonIncreasing}), $g_{\mathsf{IT}}(\eps) = 0$ if and only if $\eps \leq \calP_{\mathsf{IT}}(t^*)$.

The privacy-utility trade-off for $\calP_{\mathsf{ET}}$ can be handled similarly by replacing $\calP_{\mathsf{IT}}(t) \leq \epsilon$ with $\calP_{\mathsf{ET}}(t) \geq \epsilon$, as two highly correlated random variables posses a high mutual information but, at the same time, a small MMSE.

\section{Numerical Computation of $\calU$}
\label{Section:NumericalComputation}

Throughout this section $c\in(0,\infty)$, $r\in(0,1)$, and $s>0$ are fixed. For $t\geq0$, we let $G_t$ be the Cauchy transform of $F_t = F^{c,H_t}$ with $H_t$ defined as in \eqref{eq:Ht}. By Thm.~\ref{Theorem:MarchenkoPastur}, for each $z\in\C^+$, $G_t(z)$ is a solution to the equation
\eq{G = \frac{1-r}{z-t(1-c+czG)} + \frac{r}{z-(t+s)(1-c+czG)}.}
Alternatively, $G_t(z)$ is a root of the polynomial
\eq{P_{t,z}(G) = A_{t,z} G^3 + B_{t,z} G^2 + C_{t,z} G + D_{t,z} \in \C[G],}
where $A_{z,t} := t(t+s)c^2z^2$, $B_{z,t} := a_{t,z}(t+s)cz+b_{t,z}tcz$, $C_{z,t} := rtcz+(1-r)(t+s)cz+a_{t,z}b_{t,z}$, and $D_{z,t} := ra_{t,z}+(1-r)b_{t,z}$ with $a_{t,z} = (t(1-c)-z)$ and $b_{t,z} = (t+s)(1-c)-z$. The following lemma provides a characterization of $\Supp{F_t}$.

\begin{lemma}
\label{Lemma:Delta}
Let $\Delta_t:\R\to\R$ be the (real) polynomial given by
\al{
x \mapsto & 18 A_{x,t} B_{x,t} C_{x,t} D_{x,t} - 4 B_{x,t}^3 D_{x,t}\\
& \quad + B_{x,t}^2 C_{x,t}^2 - 4 A_{x,t} C_{x,t}^3 - 27 A_{x,t}^2 D_{x,t}^2.}
Then, $\Supp{F_t}$ is the closure of $\{x\in[0,\infty) : \Delta_t(x) < 0\}$.
\end{lemma}

\begin{figure}[t]
\centering
\includegraphics[width=0.30\textwidth]{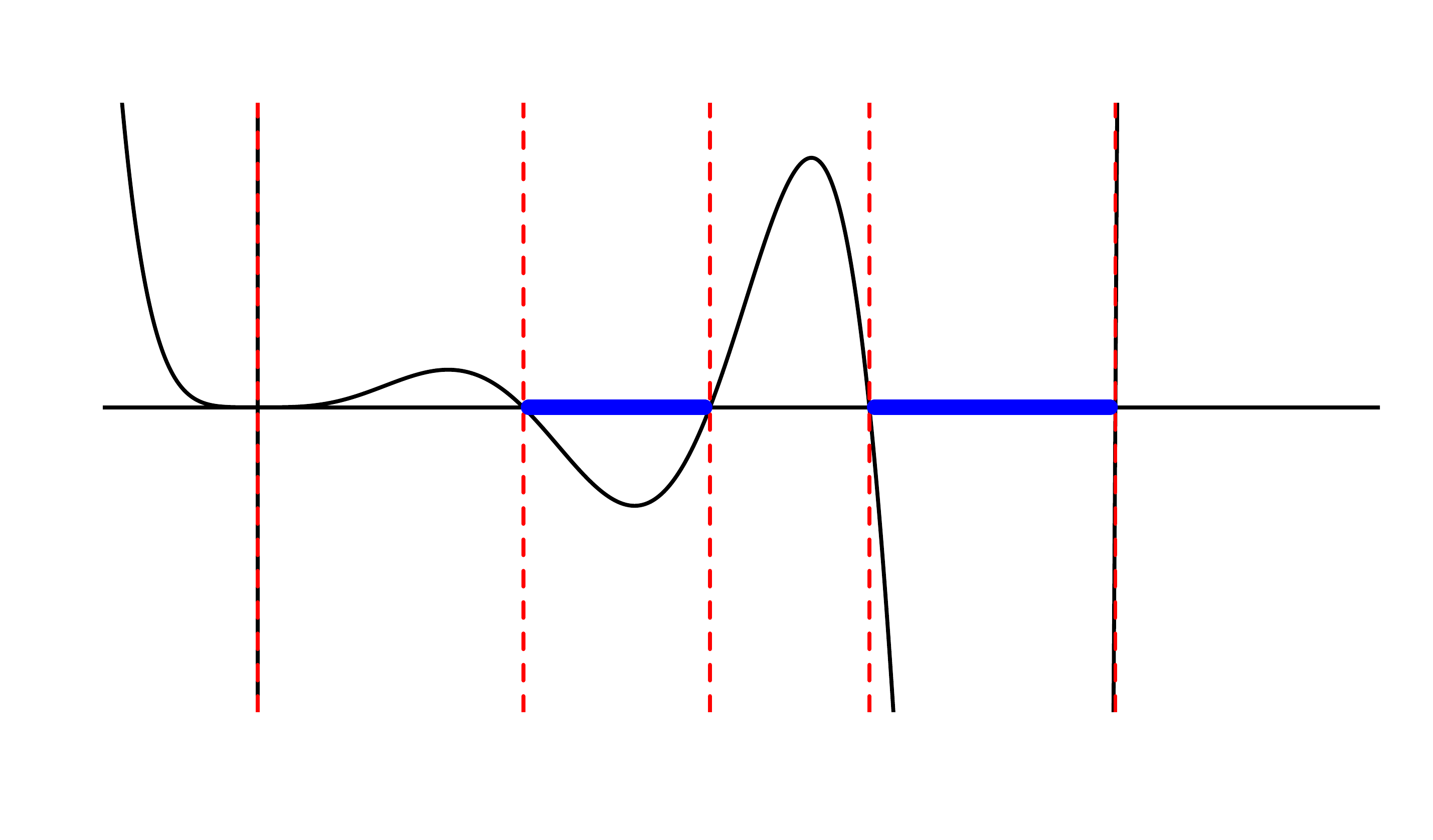}
\caption{Graph of $\Delta_t$ for $c=1/40$, $r=3/10$, $s=10$, and $t=10$.}
\label{Fig:PM}
\vspace{-10pt}
\end{figure}

The above lemma suggests a simple method to compute $\calU(t)$: find the positive roots of $\Delta_t$, identify where $\Delta_t$ is positive and negative, and subtract the roots delimiting the gap of interest. This process is depicted in Fig.~\ref{Fig:PM}, where the support of $F_t$ is represented by thick blue lines and the value of $\calU(t)$ equals the third minus the second positive root of $\Delta_t$.

\bibliographystyle{IEEEtran}
\bibliography{PPCA.bib}
\end{document}